
%
%
%
%
\magnification=1200
%
%
\hsize=31pc
\vsize=55 truepc
\hfuzz=2pt
\vfuzz=4pt
\pretolerance=5000
\tolerance=5000
\parskip=0pt plus 1pt
\parindent=16pt
%
%
\font\fourteenrm=cmr10 scaled \magstep2
\font\fourteeni=cmmi10 scaled \magstep2
\font\fourteenbf=cmbx10 scaled \magstep2
\font\fourteenit=cmti10 scaled \magstep2
\font\fourteensy=cmsy10 scaled \magstep2
\font\large=cmbx10 scaled \magstep1
%

%
%

%
%

%
%

%
%
\font\eightrm=cmr8
\font\eighti=cmmi8
\font\eightbf=cmbx8
\font\eightit=cmti8

\font\eightsy=cmsy8
\font\sixrm=cmr6
\font\sixi=cmmi6
\font\sixsy=cmsy6

\def\tenpoint{\def\rm{\fam0\tenrm}%
  \textfont0=\tenrm \scriptfont0=\sevenrm
                      \scriptscriptfont0=\fiverm
  \textfont1=\teni  \scriptfont1=\seveni
                      \scriptscriptfont1=\fivei
  \textfont2=\tensy \scriptfont2=\sevensy
                      \scriptscriptfont2=\fivesy
  \textfont3=\tenex   \scriptfont3=\tenex
                      \scriptscriptfont3=\tenex
  \textfont\itfam=\tenit  \def\it{\fam\itfam\tenit}%
  \textfont\slfam=\tensl  \def\sl{\fam\slfam\tensl}%
  \textfont\bffam=\tenbf  \scriptfont\bffam=\sevenbf
                            \scriptscriptfont\bffam=\fivebf
                            \def\bf{\fam\bffam\tenbf}%
  \normalbaselineskip=20 truept
  \setbox\strutbox=\hbox{\vrule height14pt depth6pt
width0pt}%
  \let\sc=\eightrm \normalbaselines\rm}
\def\eightpoint{\def\rm{\fam0\eightrm}%
  \textfont0=\eightrm \scriptfont0=\sixrm
                      \scriptscriptfont0=\fiverm
  \textfont1=\eighti  \scriptfont1=\sixi
                      \scriptscriptfont1=\fivei
  \textfont2=\eightsy \scriptfont2=\sixsy
                      \scriptscriptfont2=\fivesy
  \textfont3=\tenex   \scriptfont3=\tenex
                      \scriptscriptfont3=\tenex
  \textfont\itfam=\eightit  \def\it{\fam\itfam\eightit}%
  \textfont\bffam=\eightbf  \def\bf{\fam\bffam\eightbf}%
  \normalbaselineskip=16 truept
  \setbox\strutbox=\hbox{\vrule height11pt depth5pt width0pt}}
\def\fourteenpoint{\def\rm{\fam0\fourteenrm}%
  \textfont0=\fourteenrm \scriptfont0=\tenrm
                      \scriptscriptfont0=\eightrm
  \textfont1=\fourteeni  \scriptfont1=\teni
                      \scriptscriptfont1=\eighti
  \textfont2=\fourteensy \scriptfont2=\tensy
                      \scriptscriptfont2=\eightsy
  \textfont3=\tenex   \scriptfont3=\tenex
                      \scriptscriptfont3=\tenex
  \textfont\itfam=\fourteenit  \def\it{\fam\itfam\fourteenit}%
  \textfont\bffam=\fourteenbf  \scriptfont\bffam=\tenbf
                             \scriptscriptfont\bffam=\eightbf
                             \def\bf{\fam\bffam\fourteenbf}%
  \normalbaselineskip=24 truept
  \setbox\strutbox=\hbox{\vrule height17pt depth7pt width0pt}%
  \let\sc=\tenrm \normalbaselines\rm}

\def\today{\number\day\ \ifcase\month\or
  January\or February\or March\or April\or May\or June\or
  July\or August\or September\or October\or November\or
December\fi
  \space \number\year}
%
%
\newcount\secno      
\newcount\subno      
\newcount\subsubno   
\newcount\appno      
\newcount\tableno    
\newcount\figureno   
\normalbaselineskip=20 truept
\baselineskip=20 truept
%
%
\def\title#1
   {\vglue1truein
   {\baselineskip=24 truept
    \pretolerance=10000
    \raggedright
    \noindent \fourteenpoint\bf #1\par}
    \vskip1truein minus36pt}
%
%
\def\author#1
  {{\pretolerance=10000
    \raggedright
    \noindent {\large #1}\par}}
%
%
\def\address#1
   {\bigskip
    \noindent \rm #1\par}
%
%
\def\shorttitle#1
   {\vfill
    \noindent \rm Short title: {\sl #1}\par
    \medskip}
%
%
\def\pacs#1
   {\noindent \rm PACS number(s): #1\par
    \medskip}
%
%
\def\jnl#1
   {\noindent \rm Submitted to: {\sl #1}\par
    \medskip}
%
%
\def\date
   {\noindent Date: \today\par
    \medskip}
%
%
\def\beginabstract
   {\vfill\eject
    \noindent {\bf Abstract. }\rm}
%
%
\def\keyword#1
   {\bigskip
    \noindent {\bf Keyword abstract: }\rm#1}
%
%
\def\endabstract
   {\par
    \vfill\eject}
%
%
%

%
%
\def\entry#1#2#3
   {\noindent
    \hangindent=20pt
    \hangafter=1
    \hbox to20pt{#1 \hss}#2\hfill #3\par}
%
%
\def\subentry#1#2#3
   {\noindent
    \hangindent=40pt
    \hangafter=1
    \hskip20pt\hbox to20pt{#1 \hss}#2\hfill #3\par}
%
%
\def\section#1
   {\vskip0pt plus.1\vsize\penalty-250
    \vskip0pt plus-.1\vsize\vskip24pt plus12pt minus6pt
    \subno=0 \subsubno=0
    \global\advance\secno by 1
    \noindent {\bf \the\secno. #1\par}
    \bigskip
    \noindent}
%
%
\def\subsection#1
   {\vskip-\lastskip
    \vskip24pt plus12pt minus6pt
    \bigbreak
    \global\advance\subno by 1
    \subsubno=0
    \noindent {\sl \the\secno.\the\subno. #1\par}
    \nobreak
    \medskip
    \noindent}
%
%
\def\subsubsection#1
   {\vskip-\lastskip
    \vskip20pt plus6pt minus6pt
    \bigbreak
    \global\advance\subsubno by 1
    \noindent {\sl \the\secno.\the\subno.\the\subsubno. #1}\null. }
%
%
\def\appendix#1
   {\vskip0pt plus.1\vsize\penalty-250
    \vskip0pt plus-.1\vsize\vskip24pt plus12pt minus6pt
    \subno=0
    \global\advance\appno by 1
    \noindent {\bf Appendix \the\appno. #1\par}
    \bigskip
    \noindent}
%
%
\def\subappendix#1
   {\vskip-\lastskip
    \vskip36pt plus12pt minus12pt
    \bigbreak
    \global\advance\subno by 1
    \noindent {\sl \the\appno.\the\subno. #1\par}
    \nobreak
    \medskip
    \noindent}
%
%

%
%

%
%
\def\tabcaption#1
   {\global\advance\tableno by 1
    \noindent {\bf Table \the\tableno.} \rm#1\par
    \bigskip}
%
%
\def\figures
   {\vfill\eject
    \noindent {\bf Figure captions\par}
    \bigskip}
%
%
\def\figcaption#1
   {\global\advance\figureno by 1
    \noindent {\bf Figure \the\figureno.} \rm#1\par
    \bigskip}
%
%
\def\references
     {\vfill\eject
     {\noindent \bf References\par}
      \parindent=0pt
      \bigskip}
%
%
\def\refjl#1#2#3#4
   {\hangindent=16pt
    \hangafter=1
    \rm #1
   {\frenchspacing\sl #2
    \bf #3}
    #4\par}
%
%
\def\refbk#1#2#3
   {\hangindent=16pt
    \hangafter=1
    \rm #1
   {\frenchspacing\sl #2}
    #3\par}
%
%
\def\numrefjl#1#2#3#4#5
   {\parindent=40pt
    \hang
    \noindent
    \rm {\hbox to 30truept{\hss #1\quad}}#2
   {\frenchspacing\sl #3\/
    \bf #4}
    #5\par\parindent=16pt}
%
%
\def\numrefbk#1#2#3#4
   {\parindent=40pt
    \hang
    \noindent
    \rm {\hbox to 30truept{\hss #1\quad}}#2
   {\frenchspacing\sl #3\/}
    #4\par\parindent=16pt}
%
%

%
%
\def\frac#1#2{{#1 \over #2}}
%
%

%
%
\def\d{{\rm d}}
%
%
\def\e{{\rm e}}
%
%
\def\i{\ifmmode{\rm i}\else\char"10\fi}
%
%

%
%

%
%

%
%
\def\etal{{\sl et al\/}\ }
%
%

\catcode`\@=11
\def\vfootnote#1{\insert\footins\bgroup
    \interlinepenalty=\interfootnotelinepenalty
    \splittopskip=\ht\strutbox 
    \splitmaxdepth=\dp\strutbox \floatingpenalty=20000
    \leftskip=0pt \rightskip=0pt \spaceskip=0pt \xspaceskip=0pt
    \noindent\eightpoint\rm #1\ \ignorespaces\footstrut\futurelet\next\fo@t}
%
%
%
\def\ind{\hbox to 5pc{}}
%
%
\def\eq(#1){\hfill\llap{(#1)}}
%
%
\def\eql{\llap{\phantom{x}=\phantom{y}}}
%
%
\def\deqn#1{\displ@y\halign{\hbox to \displaywidth
    {$\@lign\displaystyle##\hfil$}\crcr #1\crcr}}
%
%
\def\indeqn#1{\displ@y\halign{\hbox to \displaywidth
    {$\ind\@lign\displaystyle##\hfil$}\crcr #1\crcr}}
%
%
\def\indalign#1{\displ@y \tabskip=0pt
  \halign to\displaywidth{\ind$\@lign\displaystyle{##}$\tabskip=0pt
    &$\@lign\displaystyle{{}##}$\hfill\tabskip=\centering
    &\llap{$\@lign##$}\tabskip=0pt\crcr
    #1\crcr}}
\catcode`\@=12
%
%



%
%

\def\JPA{J. Phys. A: Math. Gen.}


%
%

%
%
\endinput

\input preprint.sty
\input mssymb
\def\simle{\rlap{\raise 2pt \hbox{$<$}}{\lower 2pt \hbox{$\sim$}}}
\title{Poincar\'e maps of Duffing--type oscillators and their
reduction to circle maps}
\noindent
\underbar{{\large{I. Analytic results}}}\hfill \break
\vskip0.5cm
\author{G Eilenberger$^{\rm \dag}$ \ and K Schmidt$^{\rm \ddag}$}
\address{$^{\rm \dag}$ Institut f\"ur Festk\"orperforschung,
Forschungszentrum J\"ulich, \hfill \break
Postfach 1913, W--5170 J\"ulich, FRG}
\address{$^{\rm \ddag}$
EDS Deutschland, Eisenstra\ss{e} 56, W--6090 R\"usselsheim, FRG}

\shorttitle{Poincar\'e maps of Duffing oscillators}

\pacs{3.20.+i}

\jnl{\JPA}
\vskip0.3cm
\noindent
\vskip0.3cm
\date

\beginabstract
Bifurcation diagrams and plots of Lyapunov exponents in the
$r$--$\Omega$ --plane for Duffing--type oscillators
$$\ddot x +2r\dot x +V'(x,\Omega t) =0$$
exhibit a regular pattern of repeating selfsimilar ``tongues'' with
complex internal structure. We demonstrate here that this behaviour
is easily understood qualitatively and quantitatively from the
Poincar\'e map of the system in action--angle variables. This map
approaches the {\it one dimensional} form
$$\varphi_{n+1} = A + C \e^{-r T} \cos \varphi_n, \ \ T= \pi / \Omega$$
provided $\e^{-r T}$ (but not necessarily $C \e^{- r T}$), $r$ and $\Omega$ are
small.
We derive asymptotic (for $r$, $\Omega$ small) formulae for $A$ and $C$
for a special class of potentials $V$. We argue that these special cases
contain all the information needed to treat the general case of potentials
which obey $V'' \ge 0$ at all times.

The essential tools of the derivation are the use of action--angle
variables, the adiabatic approximation and the introduction of a
nonoscillating reference solution of Duffing's equation, with respect to
which the action-angle variables have to be determined.
These allow the explicit construction of the Poincar\'e map in powers of
$\e^{-rT}$. To first order, we obtain the $\varphi$--map, which survives
asymptotically. To {\it second} order we obtain the two--dimensional
$I$--$\varphi$--map. In $I$--direction it contracts by a factor
$\e^{-rT}$ upon each iteration.
\endabstract

\section{Introduction}

Nonlinear oscillators and their bifurcation diagrams have been widely
considered for decades, beginning with Duffing [1]. The bifurcation
diagrams have a rather regular structure asymptotically, that is,
for driving periods $T$ much larger than the oscillators own
characteristic time and for friction coefficients $r$ small enough
such that $\exp (-rT)$ remains distinguishable from zero. This
regularity has aroused quite some interest (see, for instance,
Parlitz and Lauterborn [2] and the literature cited there), but
a global understanding for its mechanism has not been achieved so far.

We shall demonstrate here that this mechanism can be rather easily
understood and used. It applies in principle to all equations of
the type
$$\ddot x +2r\dot x +V'(x,\Omega t) = 0, \eqno(1.1)$$
where
$$V(x,\tau) = V(x,\tau +2\pi), \eqno(1.2)$$
and the following additional properties are assumed: $V$ has only
{\it one} extremum $x_0(\tau)$ at all times, which corresponds to
a stable equilibrium of the system:
$$V'(x_0(\tau), \tau) = 0 \quad {\rm and} \quad V''(x_0(\tau), \tau)
\geq 0.\eqno(1.3)$$
In general, $V''(x_0(\tau), \tau) >0$ but at discrete points in time, say
$\tau =\tau_n$ including all periodic repetitions, we assume $V''(x_0
(\tau_n), \tau_n) =0$; i.e. the momentary harmonic frequency
about the equilibrium vanishes at the times $\tau_n$.

Now, for small $r$ and $\Omega$, we can apply the adiabatic theorem. In the
variable $z=\e^{rt} x$, the system is Hamiltonian with a slow time dependence
and thus its action $I$ stays constant most of the time. In the
$x$-variable's phase space, $I(t)$ decays like $\e^{-2rt}$, which is just
the phase space contraction to be expected. Around the times $\Omega t_n
=\tau_n$, however, the adiabatic theorem fails because the momentary
 harmonic frequency becomes smaller than $\Omega$.  At that point,
the action,
which had almost decayed to zero, gets kicked up to a new starting
value $I(0)$ which depends on  the angle variable $\varphi_n =
\varphi (t_n)$ via $\e^{-rT} \cos \varphi_n$. At the same time, $\varphi$
is set essentially to zero. The increment of $\varphi$ through the next
adiabatic
period $T$ is obtained by integration of $\omega (I(t), \tau)$, which
gives a leading term $A_1T$ and a term $\sim \e^{-rT} \cos \varphi_n$
from its $I$-dependence. Thus, an angular Poincar\'e map
$$\varphi_{n+1} =A_0 + A_1T +A(r,\Omega) + C (r,\Omega) \e^{-rT} \cos
\varphi_n \eqno(1.4)$$
is obtained where the functions $A$ and $C$ are finite series of positive
powers of $(\Omega^{\eta}/r),$ $\eta < 1.$ Constant factors in $A$ and $C$
have to be determined numerically
from the parameters of
$V$. The map (1.4) has a nontrivial behavior only in the range of
parameters where $ C \e^{-rT} >1$, and there it fully explains the
bifurcation diagram of the system (1.1) in the $r$-$\Omega$ plane.
There have been previous attempts (Sato \etal [3]) to reduce
Duffing type equations to circle maps. These authors did not
deduce our map (1.4), however.

In Section~2, we shall discuss the action-angle transformation for
the system (1.1). Our aim is to derive the circle map and determine $A$ and C
for model systems of the type
$$\ddot x +2r\dot x +{\rm sign} (x) \mid x\mid^q = \mid x\mid^\ell
P(\Omega t)\eqno(1.5)$$
with (essentially) arbitrary positive exponents $q$ and $\ell$.
In Section~3, we consider the case $\ell =0$ and
$$P(\Omega t) =2\Theta (\sin \Omega t) -1 ,\eqno(1.6)$$
i.e. a driving force which switches from $+1$ to $-1$ and back at
intervals $T=\pi \Omega^{-1}$. Although this case in itself does not
belong into the class described above, it models and explains correctly
the ``kick'' mechanism.

In Section~4 we treat the cases
$$P(\tau) = {\rm sgn} (\sin \tau) \mid \sin \tau \mid^p \eqno(1.7)$$  and
$$P(\tau) = \mid \sin \tau \mid^p.\eqno(1.8)$$

We introduce a slowly varying {\it reference solution}
of equation (1.5), which has analytically accessible long time behavior.
The general solution of equation (1.5) oscillates around this reference
solution;
the oscillations can be treated explicitly by use of the adiabatic
approximation.

Section~5 contains most of the technical details of the paper. Here the
Poincar\'e map is derived to {\it second} order in $\e^{-rT}$ and
it is shown, how the $r$-$\Omega$-dependence of its parameters can be
determined analytically. Restriction to {\it first} order in
$\e^{-rT}$ then yields the map (1.4).

In Section~6,  we shall argue that the
cases treated in Sections 4 and 5
cover the general case,
which can be put together from different successive maps, if necessary,
of the type described in Section~5.

\section{Adiabatic approximation}

We derive here the application of the adiabatic theorem to systems with
friction and discuss those of its aspects which are needed in the
following Sections.

Consider the equation of motion
$$\ddot x +2r\dot x + V'(x,t) = 0,\eqno(2.1)$$
$$V(x,t) = \sum^\infty_{\nu =1} {1\over \nu} a_\nu (t) x^\nu.\eqno(2.2)$$
To apply the transformation $$z =\e^{rt} x \eqno(2.3)$$ means that we
look at the phase space trajectories derived from (2.1), which always
spiral inwards, through a magnifying glass with ever increasing strength,
such that we observe the Hamiltonian motion
$$\ddot z +W'(z,t) =0 \eqno(2.4)$$
$$\eqalignno{
W(z,t) &= \e^{2rt} V(\e^{-rt} z,t) -{1\over 2} r^2z^2 & \cr
       &= a_1(t) \e^{rt} z + {1\over 2} (a_2(t)
          - r^2)z^2 + {1\over 3} a_3(t) \e^{-rt} z^3 + \ldots
   & (2.5) \cr }$$
We note that any {\it linear} term in $V$ leads to an exponentially
increasing term in
$W$. This will invalidate the adiabatic approximation for large
$rt$, which points to the (intuitively obvious) fact that one has to
expand by $\e^{rt}$ the {\it distance from the minimum} of $V$ to obtain a
{\it useful} Hamiltonian description.

We shall therefore assume $a_1=0$ and have to take care in the following
that this condition is met. {\it This is at the heart of our derivations.}

The harmonic term from (2.1) remains without exponential factor in (2.5)
but is supplemented by $-{1\over 2} r^2 z^2$. For the intended limit
$r\to 0$, this addition is irrelevant, but for computational applications
with necessarily finite $r$ it increases the accuracy if one keeps it.

The higher ``nonlinear'' terms in the equation of motion (2.4) are
eliminated exponentially in time, which reflects the fact that the
original trajectory spirals into the harmonic region.

We consider now energies and actions with respect to $z$- and $x$-variables
with $\tau$ considered a fixed parameter:
$$F={1\over 2} \dot z^2 +W(z,\tau) ,\eqno(2.6)$$
$$E={1\over 2} \dot x^2 +r x\dot x + V(x,\tau) =
\e^{-2r\tau} F,\eqno(2.7)$$
$$J(F,\tau) ={1\over 2\pi} \oint \sqrt{2[F-W(\zeta, \tau)]} \d\zeta
,\eqno(2.8)$$
$$I(E,\tau) ={1\over 2\pi} \oint \sqrt{2[E-\tilde V(\xi, \tau)]} \d\xi
= \e^{-2r\tau} J(F,\tau) ,\eqno(2.9)$$
$$\tilde V(x,\tau) = V(x,\tau) - {1\over 2} r^2 x^2 .\eqno(2.10)$$
Again the $r$-dependent terms in $E$ and $\tilde V$ vanish for
$r\to 0$ but ought to be kept in numerical computations.
The generating function for the canonical transformation to
action--angle variables (in the following abbreviated by $aav$'s)
is well known:
$$S_1(z,J,\tau) = \int^z_{z_0} \sqrt{2[F(J,\tau) -W(\zeta, \tau)]}\;
\d\zeta.\eqno(2.11)$$
Here $F(J,\tau)$ is the inverse of $J(F,\tau)$ from (2.8); it is
unique by our convexity assumption on $V$. We obtain
$$\varphi (z,J,\tau) = \omega \int^z_{z_0} {\d\zeta \over \sqrt{F,W}} =
\omega \int^x_{x_0} {\d\xi \over \sqrt{E,\tilde V}} ,\eqno(2.12)$$
where
$$\omega\,=\,{\partial F(J,\tau) \over \partial J}\,=\,{\partial
E(I,\tau) \over \partial I}\,=\,\omega (I,\tau) .\eqno(2.13)$$
The adiabatic theorem guarantees that asymptotically, for $r$ and
$\Omega$ small, the motion of the system satisfies
$$I(t) = I(0)\;\e^{-2rt} \quad {\rm i.e.} \quad J(t) = {\rm const},
\eqno(2.14)$$     and
$$\varphi (t)= \varphi(0) + \int^t_0 \omega (I(\tau),\tau)\,
\d\tau.\eqno(2.15)$$

To carry out the canonical transformation induced by (2.11) it is
convenient to define parameters $\omega_0$ and $a_n$ through the
notation:
$$W (z,\tau) = \omega^2_0 (\tau) \left( \frac{1}{2} z^2 + \frac{1}{3}
a_3 (\tau)
z^3 + \frac{1}{4} a_4 (\tau) z^4 + \ldots \right). \eqno(2.16)$$
We obtain the formulae:
$$\eqalign{
z (\varphi)&= - R \cos \varphi + \frac{a_3}{6} R^2 (\cos 2 \varphi -3)  \cr
&- \frac{1}{8} R^3 \biggl[ (\frac{a^2_3}{6} + \frac{a_4}{4}) \cos 3 \varphi +
(\frac{11 a_3^2}{9} - \frac{3a_4}{2}) \cos \varphi \biggr] + O (R^4) \cr}
\eqno(2.17)$$
and (for the momentum $\dot{z} (\varphi) = p (\varphi)$)
$$\eqalign{{1\over \omega_0}
p (\varphi)& = R \sin \varphi -  \frac{a_3}{3} R^2 \sin 2 \varphi  \cr
& + \frac{1}{8} R^3 \biggl[ (\frac{a_3^2}{2} + \frac{3a_4}{4}) \sin 3
\varphi - (\frac{19a^2_3}{9} - \frac{3a_4}{2}) \sin \varphi \biggr]
+ O (R^4) \cr} \eqno(2.18)$$
Here, we have taken $R = \sqrt{\frac{2J}{\omega_0}}$ and defined $\varphi
= 0$ by using for the lower integration limit $z_0$ in (2.11) the {\it left}
turning point for each trajectory. In conjunction with these formulae for
$z$ and $p$ we have
$$F(J,\tau) = \omega_0 J + (- \frac{5a^2_3}{12} + \frac{3a_4}{8} ) J^2
 + \ldots \quad . \eqno(2.19)$$
The Hamiltonian in action--angle variables --- which yields the
corrections to the adiabatic approximation --- reads:
$$\deqn{
H_1(J, \varphi, \tau)  = F (J, \tau) + \frac{\partial}{\partial \tau}
S_1 (z, J, \tau) \cr
\ind \eql  F(J, \tau)- \frac{1}{2} \frac{\dot{\omega}_0}{\omega_0} J
\sin 2 \varphi \eq(2.20) \cr
\ind + \frac{1}{2} J R \biggl[ (\frac{4}{9}
\frac{\dot{\omega}_0}{\omega_0} a_3 + \frac{1}{9} \dot{a}_3) \sin 3
\varphi - (\frac{4}{3}
\frac{\dot{\omega}_0}{\omega_0} a_3
- \dot{a}_3) \sin \varphi \biggr] + O (R^4) \cr}$$
Since the angular average of $H-F$ vanishes, the true action--angle variables
deviate from their adiabatic approximations (2.14) and (2.15) {\it on the
average} only to {\it second} order in the derivatives $\dot{\omega}_0,
\dot{a}_n$.
Therefore, the correction to {\it first order} in $\dot{\omega},
\dot{a}_n$ for all
quantities $z,p,J, \varphi$ can be expressed through the adiabatic variables
alone. This well known fact enhances the range of validity of our asymptotic
expansion in $r$ and $\Omega$ considerably.

We obtain for the corrected quantities $\varphi_c$ and $J_c$:
$$\eqalign{
\varphi_c & = \varphi + \frac{1}{4} \frac{\dot{\omega}_0}{\omega^2_0}
\cos 2 \varphi \cr
& - \frac{1}{2 \omega_0} R \biggl[ ( \frac{2}{9}
\frac{\dot{\omega}_0}{\omega_0}
a_3 + \frac{1}{18} \dot{a}_3 ) \cos 3 \varphi - (2
\frac{\dot{\omega}_0}{\omega_0} a_3
- \frac{3}{2} \dot{a}_3 ) \cos \varphi \biggr] \cr} \eqno(2.21)$$
and
$$\eqalign{
J_c & =  J \biggl\{ (1 - \frac{1}{2} \frac{\dot{\omega}_0}{\omega^2_0})
\sin 2 \varphi \cr
& + \frac{1}{2 \omega} R \biggl[ (\frac{4}{9}
\frac{\dot{\omega}_0}{\omega_0}
a_3 + \frac{1}{9} \dot{a}_3 ) \sin 3 \varphi - ( \frac{4}{3}
\frac{\dot{\omega}_0}{\omega_0}
a_3 - \dot{a}_3 ) \sin \varphi \biggr] \biggr\} \cr} \eqno(2.22)$$
with relative errors in second order of time derivatives and $R^2$ times
first order of time derivatives.
Here, the quantities $J, \varphi$ on the rhs are the adiabatic ones
from (2.14), (2.15). The quantities $z$ and $p$ are then obtained by
inserting $J_c$ and $\varphi_c$ for $J$ and $\varphi$ in (2.17) and
(2.18).
We shall refer to these formulae in Section~5.

The Hamiltonian (2.20) can be processed further. This is particularly
useful, if one intends to construct numerically Poincar\'e maps in
$aav$'s for given fixed values of $r$ and $\Omega$ instead of using the
asymptotic formulae of Section~5 which contain the parameter dependence in
explicit analytic form.

The transformation
$$\sqrt{\frac{J}{\omega_0}} \cos \varphi = g \sqrt{\tilde{J}}
\cos \tilde{\varphi}, \eqno(2.23)$$
$$\sqrt{J \omega_0} \sin \varphi = \frac{1}{g} \sqrt{\tilde{J}}
(\sin \tilde{\varphi} - g \dot{g} \cos \tilde{\varphi})\,,\eqno(2.24)$$
where $g(\tau)$ obeys the differential equation
$$\ddot{g} + \omega^2_0 (\tau) g - \frac{1}{g^3} = 0 \, ,\eqno(2.25)$$
eliminates the term linear in $J$ from equations (2.20), (2.21) and
(2.22). It is generated by the function
$$S_2(\varphi, \tilde{J}, \tau) = \tilde{J}
\arctan (\omega_0 g^2  \tan \varphi + g \dot{g}), \eqno(2.26)$$
and yields the new Hamiltonian
$$H_2(\tilde{J}, \tilde{\varphi}, \tau) ={\tilde{J} \over {g^2(\tau)}}
+O(\tilde{J}^{3/2}). \eqno(2.27)$$
{}From this, the $\varphi$--dependence can be eliminated entirely through
an Ansatz for a third generating function $S_3$ in powers of
$\tilde{J}^{1/2}$. Its coefficients are to be determined recursively
from explicitly solvable linear {\it first} order differential
equation in $\tau$. The total effect of $S_1$, $S_2$ and $S_3$
could be more conveniently accomplished in one step, however, by
starting directly with an Ansatz for $z$ and $p$ of the type of
equations (2.17),(2.18), where $\omega_0$ is replaced by $g^{-2}$
throughout.

We shall later apply the formalism to potentials $V(y,t)$ which are
homogeneous functions of degree $q+1$ in $y$ and some $x_0(t)$;
more specific:
$$\eqalign{V(y,x_0)&= {1\over q+1} [(y+x_0)^{q+1} -(q+1)yx_0^q
 -x_0^{(q+1)}] \cr
&-{x_0^{(q-\ell)} \over \ell +1} [(y+x_0)^{\ell +1} -(\ell +1) yx_0^\ell
-x_0^{(\ell+1)}] \cr
&={q-\ell \over 2} x_0^{(q-1)} y^2 +{(q+\ell -1)(q-\ell) \over 6}
x_0^{(q-2)} y^3 + \ldots \cr} \eqno(2.28)$$
Using this homogeneity, we obtain the scaling relation [dropping the
${1\over 2} r^2$ term from (2.10)]
$$L^{q+1}\,E(I,x_0) = E(L^{{q+3\over 2}} I,\,Lx_0).\eqno(2.29)$$
This yields expansions
$$E(I,x_0) = \sum^\infty_{n=1} c_n\,x_0^{(q+1)-{n\over 2} (q+3)} I^n
\eqno(2.30)$$
and
$$\omega (I,x_0) = c_1 x_0^{{q-1\over 2}} +2c_2 x_0^{-2} I
+ \ldots \eqno(2.31)$$
The parameters in (2.16) are then given by
$$\omega^2_0 = (q-\ell) x_0^{q-1}, \eqno(2.32)$$
$$a_3 = \frac{1}{2} (q + \ell -1) x_0^{-1}, \eqno(2.33)$$
$$a_4 = \frac{1}{6} [q^2 + q \ell + \ell^2 - 3 (q + \ell) + 2]
x_0^{-2}; \eqno(2.34)$$
generally, we have $a_n \sim x_0^{2-n}$.

Finally:
$$c_1=(q-\ell)^{1\over 2},\quad c_2 = -{1\over 48} [2q^2 +7q\ell +
2\ell^2 -q-\ell -1].\eqno(2.35)$$

\section{Step function driving as kick mechanism}

In this Section we shall demonstrate, with the technically simplest
model, the mechanism which leads to the map. This model turns out to
correctly describe the kick mechanism encountered in Section~4.
We consider
$$\ddot x+2r\dot x + x^q = \pm 1 = 2\Theta (\sin \Omega t) -1,\eqno(3.1)$$
i.e.~the sign of the driving force switches at times $nT$. (Note that
we use for convenience of notation the symbol $T$ for the {\it half}
period $\pi /\Omega$.) We assume $q$ to be an odd integer, to avoid the
notational complication sgn$(x) \mid \! x \! \mid^q$. In the final
formulae, however, any $q>1$ may be inserted. The equation of motion
(3.1) is derived from a potential
$$V_\pm  (y) = {1\over q+1} (y\pm 1)^{q+1} \mp y -{1\over q+1}, \eqno(3.2)$$
i.e. $x = x_0 + y$ with $x_0=\pm 1$ in (2.28).
The region in parameter space to be considered is given by
$$r \ll 1 \ll T.\eqno(3.3)$$
{\it Crucial for our derivation is the further assumption}\,
$\e^{-rT} \! \ll \! 1$ as we shall systematically neglect higher powers of
this factor. This will be justified later.

The mechanism which leads to the map is then easily understood with the
help of Figure~1.
Suppose, we start a motion at $t=0$ on the right potential curve
$V_+(y)$ exactly at the point $P$, with $E_0=2$, $I_0=I(E_0)$ and
$\varphi (0)=0$. This defines $E_0, I_0$ and our convention
for $\varphi =0$. $I_0$ and $\omega_0$ are functions of $q$;
however, for simplicity of notation, we shall not make this explicit
in our formulae.

The trajectory spirals down into the harmonic region, where it reaches
some point $Q$ at time $T$ with
$I_1=I_0\,\e^{-2rT}$, and some $\varphi_1$ to which belong the quantities:
$$y_1 =-\sqrt{{2I_1\over \omega_1}} \cos \varphi_1, \quad \dot y_1 =
\sqrt{2I_1\omega_1} \sin \varphi_1 \eqno(3.4)$$
proportional to $e^{-rT}$. The harmonic frequency here is $\omega_1 =
q^{{1\over 2}}$. At that moment, the potential
switches to $V_-(y)$ and the energy is instantaneously raised to point
$R$ -- this is the ``kick''. The height of $R$ above the point $\tilde R$
on the potential curve $V_-(y)$ is the kinetic energy (the same as the
height of the point $Q$ above the curve $V_+(y)$). It is proportional to
$\dot y^2 \sim \e^{-2rT}$ and is thus negligible to lowest order in
$\e^{-rT}$. Essential for the kick is the first order term $\delta E_0 =
y_1 \cdot V'_-(2) \sim \e^{-rT}$.

The next cycle of length $T$ begins with $I(0)=I_0+\delta I_0$ and some
$\varphi(0) =\delta \varphi_0$.
For this cycle, we use the notation implied by symmetry, i.e. $\varphi
=0$ at the point $\tilde{R}$.

We have (counting the angle clockwise from the point $\tilde{R}$)
$$\delta\varphi_0 = - \omega_0 \frac{\dot{y}_1}{V'_-(2)} =
- \frac{\omega_0}{2} (2 I_0 \omega_1)^{\frac{1}{2}} \e^{-r T} \sin \varphi_1
\eqno(3.5)$$
and
$$\delta I_0 ={\partial I\over \partial E} (E_0) \cdot y_1
\cdot V'_-(2)
=-{2\over \omega_0} ({2I_0\over \omega_1})^{1\over 2}\,\e^{-rT}
\cos \varphi_1. \eqno(3.6)$$
To lowest order in $\e^{-rT}$ the $\delta \varphi_0$-term can be neglected
besides the angle
increment $\Delta \varphi$ of (3.11); i.e. the reinjection occurs
with $\varphi = 0$. ($\delta\varphi_0$ does play a role in the
general case treated in Section~5).
This cycle ends at point $S$ with $I_1+\delta I_1=\e^{-2rT}
(I_0+\delta I_0)$, where
$\delta I_1$ is smaller than $I_1$ by a factor $\e^{-rT}$ and thus again
negligible. {\it This is the crucial step} which eliminates one phase
space dimension and leads to a map in $\varphi$ alone.

It means that in the full Poincar\'e map in $I$-$\varphi$ coordinates the
contraction towards to line $I=I_1$ is so strong that this variable
may be neglected altogether.
The ``kick''-mechanism is thus based on a discontinuous change of the
variable $y$ which measures the position of the oscillator with respect
to the minimum of the potential; the velocity $\dot y$ at that moment
is here irrelevant, as it only leads to contributions of higher order in
$\e^{-rT}$. This mechanism is our ``model $A$''.
This alternation between kicks and adiabatic motion (which, as we shall
see in Section~4, also describes the general case) seems to lie at the
base of the ``flip and twist map'' described by Brown and Chua [4].

A different kick mechanism, ``model $B$'', will also be encountered in
Section~4.
In that model, the roles of $y$ and $\dot{y}$ are interchanged; $\dot{y}$
is suddenly increased to $\dot{y} + 2 p_0$ while $y$ remains unchanged.
We obtain
expressions analogous to those of model A.

Returning to model $A$, we shall show below that the increment in $\varphi$
during one cycle is of the form
$$\Delta \varphi =\omega_1 T +{1\over r} A_2 -{\omega_1 -\omega_0 \over
2I_0} {1\over r} \delta I_0,\eqno(3.7)$$
to linear order in $\delta I_0$.
This, together with (3.6), yields the map
$$\varphi_{n+1} =\omega_1 T+{1\over r} (A_2+B\,\e^{-rT} \cos \varphi_n),
\eqno(3.8)$$
where the constants $A_2$ and $B$ have to be determined numerically.
They contain all the relevant information on the nonlinearity of the
system.

The map (3.8) can be written
$$\varphi_{n+1} = \alpha + \beta \cos \varphi_n, \quad \alpha =
\omega_1 T+{1\over r} A_2, \quad \beta = {B\over r} \e^{-rT}. \eqno(3.9)$$
It obviously yields $2\pi$ periodicity of the bifurcation diagram in
the $\alpha$--$\beta$ parameter plane. In the $r$--$\Omega$--plane, the loci
of equal features (constant $\beta =B\e^{-K}$) lie on the curves
$$\Omega = \pi r (K-log r)^{-1} .\eqno(3.10)$$
$\Omega$ decreases faster than $r$, albeit only logarithmically.
Put differently, $\e^{-rT} \sim r$ for the parameter range of interest;
this justifies the neglect of higher orders of $\e^{-rT}$.

We shall now complete the derivation of the relation (3.7). In
$$\Delta \varphi = \varphi(T) -\varphi(0) =\omega_1 T+\int^T_0 (\omega(I)
-\omega_1)\,\d t,\eqno(3.11)$$
we use from the expansion (2.31)
$$Q(I) = {\omega(I) -\omega_1 \over I} =2c_2+3c_3 I+\ldots \eqno(3.12)$$
and $\d t =-{1\over 2r} {\d I\over I}$ to obtain
$$\varphi(T) -\varphi(0) =\omega_1 T +{1\over 2r} \int^{I(0)}_{I(T)}
Q(I)\,\d I.\eqno(3.13)$$
For this integration, we replace the lower bound by 0 since $Q$ behaves
regularly at 0 and $I(T)$ is negligible. We expand the upper bound to
first order in $\delta I_0$ to obtain (3.8)
$$\Delta \varphi =\omega_1 T+{1\over 2r} \int^{I_0}_0 Q(I) \d I -
{\omega_1 -\omega_0 \over 2I_0} \cdot {1\over r}\,\delta I_0 .\eqno(3.14)$$

\section{The models $P(\tau) =\sin^p \tau$}

We consider the system
$$\ddot x+2r\dot x +x^q =x^\ell (\sin \Omega t)^p.\eqno(4.1)$$
This contains all information to understand the general case mentioned
in Section~1, as we shall argue in Section~6. We take $q$ an odd, $\ell$
an even, and $p$ an arbitrary positive integer only to avoid notational
complications, our results being valid for all real values
$$q>1,\quad \ell = 0,1 \ {\rm or} \ \ge 2, \quad (q-\ell) >2p>0 .
\eqno(4.2)$$
The necessity of the third inequality will become clear below (5.20).
Arbitrary exponents are meant to imply for the driving force
$${\rm sgn}(\sin \Omega t) \mid \sin \Omega t\mid^p \cdot
\mid x\mid^\ell \qquad {\rm  model}\,A \eqno(4.3)$$   or
$$\mid \sin \Omega t\mid^p \cdot \mid x\mid^\ell \qquad {\rm model}\,B
\eqno(4.4)$$
and, as already mentioned, ${\rm sgn}(x)\cdot\mid x\mid^q$ for the
anharmonic force. \hfill \break
For convenience, we introduce the exponent
$$\delta ={p\over (q-\ell)} < {1\over 2}.\eqno(4.5)$$
As mentioned in Section~2, the adiabatic approximation requires the
introduction of a new variable $y(t)$ via
$$x(t) =x_0(t)+y(t),\eqno(4.6)$$
such that the total potential $V$ for the $y$--motion has its minimum
at $y=0$ for all times. The ``naive'' choice for $x_0(t)$, we call it
$x_a(t)$ (as it is a very good approximation for most of the time),
would be
$$x_a(t) = (\sin \Omega t)^\delta.\eqno(4.7)$$
However, its insertion into (4.1) via (4.6) yields, of course, additional
terms from the derivatives of $x_a$ which, though of higher order in
$\Omega$, diverge for $t\to 0$ and thus spoil the desired property of
$V$.

The equation of motion in $y(t)$ reads
$$\eqalign{&
\ddot y+2r\dot y+(y+x_0)^q -x_0^q -[(y+x_0)^\ell -x_0^\ell]
\sin^p \Omega t = \cr
& -[\ddot x_0 +2r\dot x_0 +x_0^q -x_0^\ell
\sin^p \Omega t] . \cr} \eqno(4.8)$$
We require the rhs to vanish, thus $x_0$ itself must be a solution of
(4.1), the ``reference solution''. For the application of the adiabatic
approximation to the motion described by the lhs of (4.8), $x_0(t)$
must be a {\it special} solution, namely a creeping
solution which only varies on some time scale $\Omega^{-\eta}
\gg 1$, and {\it not} on time scale 1, as the general solution
does, i.e.~{\it it must not oscillate}.

Such a solution indeed exists during one half period $T$. It can be
defined through an asymptotic expansion in $r$ and $\Omega$.
It will be close
to $x_a$ except near $t=0$ and $t=T$. Its qualitative phase portrait
(and the symmetric one for the next half period of model $A$) is shown
in Figure~2.

\noindent These reference solutions {\it do not join together} at $t=T$,
as they do in the harmonic case.
Instead, for model $A$, it jumps from $x_{00}$ to $-x_{00}$ with fixed
$\dot x_0$, whereas for model $B$, $x_0$ remains unaltered and $\dot x_0$
jumps from $-p_{00}$ to $+p_{00}$. {\it This provides exactly for the two
kick mechanisms discussed in the previous Section!}

During each half period $T$, the potential $V(y,x_0(t),t)$ varies slowly
in time and thus allows for the adiabatic approximation. At $t=nT$, the
action $I$ is always kicked up again.

The determination of the kick parameter requires further investigation
on $x_0(t)$. This we shall consider next.

To investigate $x_0(t)$ in the vicinity of $t=0$, it suffices to
consider the equation of motion
$$\ddot x+2r\dot x +x^q =x^\ell (\Omega t)^p .\eqno(4.9)$$
We rescale the variables as
$$x(t) = \Omega^\gamma \xi (\tau) ,\quad t=\Omega^{-\eta} \tau
\eqno(4.10)$$      with the exponents
$$\gamma ={2\delta \over 2+(q-1)\delta}, \quad \delta ={p\over q-\ell},
\quad \eta ={(q-1)\delta \over 2+(q-1)\delta},\eqno(4.11)$$
to obtain
$$\ddot \xi +2\rho \dot \xi + \xi^q =\xi^\ell \tau^p ,\eqno(4.12)$$
$$\rho =r\Omega^{-\eta} .\eqno(4.13)$$
The smooth solution $\xi_0 (\tau)$ of (4.12), which then corresponds to
 the desired
function $x_0(t)$ near $t=0$, can be approximated by an asymptotic
series, which is obtained by iteration of
$$\xi_{n+1} (\tau) = [\tau^p -\xi_n^{-\ell} (\ddot \xi_n +2\rho
\dot \xi_n)]^{{1\over q-\ell}} \eqno(4.14)$$
starting with an initial function $\xi_1 (\tau) =\tau^\delta$. The
series has the form
$$\xi_0 (\tau) =\tau^\delta (1+\sum^\infty_{n=1} P_n(\rho \tau)
\tau^{-n\lambda}), \qquad \lambda = 2+(q-1) \delta\,, \eqno(4.15)$$
where the $P_n$ are certain polynomials of order $n$.
It definitely diverges as $P_n(0)$ grows roughly like
$(n!)^2$. Nevertheless, it is an asymptotic approximation valid for
large $\tau$ and serves several important purposes:
\item{a)} it yields reliable initial conditions at large $\tau$ for the
determination of $\xi_0 (\tau)$ by numerical integration of the
equation of motion (4.12);
\item{b)} it shows that asymptotically, for small $\rho$, the
$\rho$-dependence of the required solution is negligible.
{}From (4.13) and (3.10), we have $\rho \approx \Omega^{1-\eta} \ll 1$.
{\it The reference solution $x_0(t)$ can thus
be determined with $r=0$!}\ This can be made intuitive in the
following manner (Figure~3).
\itemcon The minimum of the total potential in (4.1)
{\it decelerates} during the first quarter period. In order for
$x_0(t)$ to follow this deceleration without oscillations, the
motion must start with a very particular initial velocity towards the right
and a very particular position on the rhs of the minimum (see Figure 3).
During some initial time interval of order $\Omega^{-\eta}$, the
required relative deceleration $\ddot x/\dot x$ is much larger than the
friction coefficient $r$. On the other hand, at the time $\tau \approx
{1\over r}$, when the friction effect is being felt, $x_0(t)$ has
already approached $x_a(t)$ very well and the velocity is small.

\itemcon The same consideration also demonstrates that the trajectories
$x_0(t)$ of Figure~2 remain in their respective half planes, i.e.~they
do not cross the $\dot x$-axis.
\item{c)} The solution (4.15) shows that $\xi_0 (\tau)$ approaches
$\tau^\delta$ algebraically. Described in the original variables, this
happens on a time scale $\Omega^{-\eta}$. Thus, we have fully
separated time scales asymptotically for $\Omega \to 0$: $x_0(t)$
approaches $(\Omega t)^\delta$ on the scale $\Omega^{-\eta}$ whereas
the difference between $\Omega t$ and  $\sin \Omega t$ is only being
felt on the scale $\Omega^{-1} \gg \Omega^{-\eta} \gg 1$.

We arrived at the following description for $x_0(t)$ in the upper right
quarter of the phase plane of Figure~2 (the other quarters are
obtained by symmetry):
$$x_0(t) = (\sin \Omega t)^\delta +\Delta x_0 (t) \eqno(4.16)$$
where
$$\Delta x_0(t) =\Omega^\gamma (\xi_0 (\tau) -\tau^\delta), \quad
\tau =\Omega^\eta t,\eqno(4.17)$$
and $\xi_0 (\tau)$ is the unique slowly varying solution of
$$\ddot \xi_0 +\xi_0^q =\xi_0^\ell \tau^p.\eqno(4.18)$$
This reference solution has similarity with and is related to asymptotic
solutions of Duffing's equations considered by Byatt-Smith [1].

In particular, we define
$$\xi_1 = \xi_0 (0), \quad \xi_2 = \dot{\xi}_0 (0). \eqno(4.19)$$
$\xi_1$ and $\xi_2$, like $I_0$ and $\omega_0$ from Section~3, belong to a set
of
about a dozen ``universal'' numbers, that have to be determined
numerically to obtain all the prefactors of the map. By ``universal''
we mean independent of $r$ and $\Omega$. They do depend on the
exponents $\ell, p$ and $q$ and on whether one considers model $A$
or $B$.

To calculate creeping solutions for finite values of $\Omega$ and $r$
numerically, one generates initial conditions for the differential
equation at $t=T/2$ by entering with the series
$$x_0(t)=\sin^\delta \Omega t (1+\sum^\infty_{n=1} R_n \ \Omega^{-2n}
\sin^{-n\lambda} \Omega t) \eqno(4.20)$$
into the equation (4.1). The coefficients $R_n$ depend on $\cos \Omega t$,
$\sin \Omega t$ and $r/\Omega$. The series is asymptotic for
$\Omega \to 0$ and yields useful initial conditions at $t=T/2$
for creeping solutions if $\Omega^2 \lesssim 10^{-1}$.

\section{Determination of the map parameters}
In this Section we shall derive the angular Poincar\'e map
$$\varphi_{n+1} = \alpha + \beta \cos \varphi_n \eqno(5.1)$$
for systems of type (4.1) and give expressions for its parameters $\alpha$
and $\beta$ in terms of $r$ and $\Omega$, valid asymptotically for small
$r$ and $\Omega$.
We shall obtain representations
$$\alpha = A_0 + A_1 T + \sum_{\nu =2}^N A_\nu F_\nu (r, \Omega)
\eqno(5.2)$$
$$\beta = \e^{-r T} \biggl[ C + B_1 \cdot \sum_{\nu =2}^N (\nu -1)
 A_\nu F_\nu (r, \Omega) \biggr].\eqno(5.3)$$
Here, the constants $A_\nu, B_1$ and $C$ will be expressed in terms of
the basic set
of numbers mentioned earlier. The functions $F_{\nu}$ are proportional
to positive powers
of $\Omega^{\eta}/r$ when $\Omega \ll r$, but we need not make this
additional assumption.
The formulae (5.2), (5.3) are asymptotic in the sense that all
terms are systematically neglected, which are smaller by factors
$\e^{- r T}$ or $(\Omega^{\eta}/r)^{-c}$ compared to what was kept.
In the following, we shall derive our results by first tracing crudely
the generation of the map and later fill in details.

Suppose, for the moment being, that the adiabatic approximation for
the motion with respect to the reference solutions were valid
rigorously at all times. At the end of the $(n-1)$--th cycle the
system coordinates are close to the endpoint of the reference
solution, denoted by $2T$ on the lhs of Figure 2.

The $n$--th cycle from $t=0$ to $t=T$ then starts with initial
$aav's$, which are conveniently written as:
$$I_n(0) = \Omega^{2\gamma +\eta} (I_a+\delta I_n) =
\Omega^{2\gamma +\eta} I_a (1+e^{-rT} i_n), \eqno(5.4)$$
$$\varphi_n(0) = \varphi_a +\delta \varphi_n .\eqno(5.5)$$

Here, we have denoted as $I_a$ and $\varphi_a$ those values which
obtain if we start {\it exactly} at the endpoint $2T$ of the lhs
reference solution in Figure~2. The scaling factor
$\Omega^{2\gamma +\eta}$ transforms $I$ into the $\xi$,$\tau$--scale
of (4.10) such that $I_a$ becomes asymptotically independent
of $\Omega$.
Continuing to $t=T$, $I_n$ and $\varphi_n$ evolve into
$$I_n=I_n(T)=e^{-2rT} I_n(0), \quad {\rm and\,some} \quad \varphi_n =
\varphi_n(T) \eqno(5.6)$$
derived below. Near $T$ the motion is close to harmonic as in the
model of Section 3; its coordinates and velocities are then given
by linear combinations of the quantities
$$u_1=e^{-rT} (1+e^{-rT} i_n)^{1/2} e^{i\varphi_n} \quad {\rm and}
\quad u_2=u_1^*\,, \eqno(5.7)$$
with factors $(I_a/2\omega_0)^{1/2}$ and scaling powers of $\Omega$,
the latter depending on whether one uses the $x,t$ or the
$\xi, \tau$--scaling. Performing a specular reflection in the
$x,\dot x$ plane (considering model $A$), the system is again near
the point $2T$ of Figure 2. More specifically, it is apparent from
the analogous discussion in Section 3
that its deviation from the endpoint of the reference solution is
of order $e^{-rT}$. The initial conditions for the $(n+1)$st cycle
can now be expanded in powers of the small quantities $u_\nu$ and
we can write to {\it second}  order (using summation convention)
$$e^{-rT} i_{n+1} = K_\nu u_\nu +{1\over 2} K_{\nu \mu} u_\nu u_\mu \,,
\eqno(5.8)$$  $$\delta \varphi_{n+1} =P_\nu u_\nu +{1\over 2}
P_{\nu \mu} u_\nu u_\mu \, . \eqno(5.9)$$
Since $I$ and $\varphi$ are real, these coefficients are more
conveniently expressed by amplitudes and phases as
$$K_1=K_2^*={1\over 2} B_1 e^{i\chi_1}, \quad P_1 = P_2^* =
{1\over 2} C_1 e^{i\psi_1} ,\eqno(5.10)$$
$$K_{11}=K_{22}^*={1\over 2} B_2 e^{i\chi_2}, \quad P_{11}= P_{22}^* =
{1\over 2} C_2 e^{i\psi_2} ,\eqno(5.11)$$
$$K_{12}=K_{21}^*={1\over 2} B_3, \quad P_{12} = P_{21}^* =
{1\over 2} C_3\;, {\rm real} .\eqno(5.12)$$
Also, the transformation from $u_\nu$ to $I, \varphi$ is canonical;
this requires
$$2{\rm Im}(K_1\cdot P_2) = {1\over 2} B_1\,C_1 \sin(\chi_1 -\psi_1)
=1 \eqno(5.13)$$      and
$$K_{11}P_2 +K_1P_{12} = P_{11}K_2+P_1K_{12} .\eqno(5.14)$$
Together with $I_a$ and $\varphi_a$ we have thus twelve numerical
constants which determine the map to second order in $e^{-rT}$.
We shall show below that these constants are asymptotically
independent of $\Omega$ and $r$ and thus ``universal''; they do depend
on $p$, $q$ and $\ell$, of course.

The map is thus a systematic expansion in powers of $e^{-rT}$ and we
obtain to second order in $e^{-rT}$ for the action
$$i_{n+1} = (1 + \frac{1}{2} \e^{-rT} i_n) B_1 \cos (\varphi_n +\chi_1) +
\e^{-rT} \biggl[ B_2 \cos (2 \varphi_n + \chi_2) + B_3 \biggr]
\eqno(5.15)$$
and for the angle
$$\eqalign{
\varphi_{n+1}(0)&=\varphi_a \cr
\phantom{=} & +e^{-rT} (1+{1\over 2} e^{-rT} i_n)
C_1 \cos (\varphi_n +\psi_1) \cr
\phantom{=} & +e^{-2rT} (C_2 \cos (2\varphi_n +\psi_2)
+C_3).\cr} \eqno(5.16)$$
To complete the map, we need the evolution of $\varphi_{n+1}(0)$ to
$\varphi_{n+1}(T)$. We have
$$\varphi_{n+1}(T) =\varphi_{n+1}(0) + \int^T_0 \omega_a (I_{n+1} (t),
t)\,dt \quad . \eqno(5.17)$$
Except for $t$--values close to $0$ and $T$,
the function $x_0 (t)$ is well approximated by
$x_a (t) \! = \! (\sin \Omega t)^{\delta}$.
Consequently, the expansion (2.31) is valid with $x_0=x_a (t)$. We
write
$$\omega_a (I (t), t) = \sum_{\nu =1}^{\infty} \nu  c_\nu I^{\nu -1} (t)
 \cdot (\sin \Omega t)^{-2 \delta_\nu } \eqno(5.18)$$
where we introduced
$$\delta_\nu  = \frac{\delta}{2} [\frac{\nu }{2} (q+3) - (q+1)] ,
\quad \kappa_\nu  = 1 - 2 \delta_\nu .\eqno(5.19)$$
We rewrite this, using $\gamma = \delta (1 - \eta)$:
$$\omega_a = c_1 \sin^{-2 \delta_1} (\Omega t) + \sum_{\nu \ge 2} \nu
c_\nu (I_a + \delta I)^{\nu -1} \Omega^{\eta \kappa_\nu }
(\frac{\sin \Omega t}{\Omega})^{-2\delta_\nu } \e^{-2(\nu -1) rt}.
 \eqno(5.20)$$
Consider this expression in the extreme asymptotic limit $1 \gg r \gg \Omega$
where we may linearize $\sin \Omega t$ in the region where
$\e^{- 2rt }$ is nonnegligible. The $\nu$--th term in the sum (5.20)
contributes a term to the angle increment $\int \omega_a dt$ proportional to
$$F_{\nu, asymptotic} = \Omega^{\eta \kappa_\nu} \int_0^{\infty}
\tau^{-2\delta_\nu } \e^{-2(\nu -1) r\tau} \d \tau = \frac{\Gamma
(\kappa_\nu )}{(\nu -1)^{\kappa_\nu }}
\left( \frac{\Omega^{\eta}}{2 r} \right)^{\kappa_\nu }. \eqno(5.21)$$
We are interested only in
terms with {\it positive} $\kappa_\nu $; any term with negative
 $\kappa_\nu $ yielded an
asymptotically vanishing contribution from the upper integration limit
and diverged at $\tau = 0$. Consequently, we sum the terms in (5.18)
only up to $\kappa_{max} = N$, such that $\kappa_N > 0$,
 $\kappa_{N+1} < 0$. (We do not
consider the rather special case that some $\kappa_N = 0)$.

The condition $\delta < \frac{1}{2}$
of (4.4) guarantees $N \ge 2$, otherwise no $I$-dependence would survive
in (5.20) and thus $\beta$-values of interest could not be realised
with $r \ll 1$, \  $\Omega \ll 1$ {\it and} $\e^{- r T} \ll 1$
simultaneously.

For {\it finite} $r$ and $\Omega$ we obtain from the adiabatic
approximation
$$\varphi_{n+1}(T) -\varphi_{n+1}(0) =
\int_0^T \omega_a \d t = A_1 T + \sum_{\nu = 2}^N
\nu  c_\nu  (I_a + \delta I)^{\nu -1} F_\nu (r, \Omega) \eqno(5.22)$$
with
$$A_1 = \frac{c_1}{\sqrt{\pi}} \frac{\Gamma (\frac{\kappa_1}{2})}{\Gamma
(\frac{\kappa_1 + 1}{2})} \eqno(5.23)$$
and
$$F_\nu (r,\Omega) =\Omega^{(\eta -1) \kappa_\nu} F(\delta_\nu,
 (\nu -1) {r\over \Omega}) \quad , \eqno(524)$$
where $F$ is given by
$$\eqalign{F(\delta, \mu) &= \int^\pi_0 e^{-2\mu t} (\sin t)^{-2\delta}
dt \cr
&= 2^{-\kappa} e^{-\pi \mu} {2\pi \Gamma (\kappa) \over \Gamma
(1-\delta +i\mu) \Gamma (1-\delta -i\mu)} . \cr} \eqno(5.25)$$
We could not find this integral in tables and arrived at equation (5.25)
through analytical continuation to $\beta =e^{-i{\pi \over 2}}$ of [5]:
$$\int_0^{\infty} \e^{-2 \mu t}(\frac{1}{\beta} \sinh \beta t)^{-2 \delta}
\d t= (2\beta)^{-\kappa} \Gamma (\kappa) \frac{\Gamma (\delta +
\frac{\mu}{\beta})} {\Gamma (1-\delta + \frac{\mu}{\beta})}. \eqno(5.26)$$

Collecting the terms from equations (5.5), (5.9), (5.15), (5.16)
and (5.20) we obtain for the $\varphi$--map --- to second order in
$e^{-rT}$ :
$$\eqalign{ \varphi_{n+1} &= \varphi_{n+1}(0) + \int^T_0
\omega_a (I_{n+1} (t),t) dt \cr
&= \varphi_a +\left( 1+{1\over 2} e^{-rT} i_n\right) C_1 \cos (\varphi_n
+\psi_1) \cr
&\, +e^{-rT} \bigl[ C_2 \cos (2\varphi_n +\psi_2) +C_3 \bigr] \cr
&\, +A_1T + \sum^N_{\nu =2} A_\nu F_\nu (r,\Omega) (1+e^{-rT}
i_{n+1})^{\nu -1} \cr} \eqno(5.27)$$     with
$$A_\nu = \nu c_\nu I_a^{\nu -1} \;, \quad (\nu \geq 2) .\eqno(5.28)$$
Sorting the terms of $\varphi_{n+1}$ in increasing powers of
$e^{-rT}$, we obtain
$$\varphi_{n+1} = O_0 + \e^{-rT} O_1 + \e^{-2rT} O_2, \eqno(5.29)$$
where
$$O_0 = \varphi_a + A_1 T + \sum^N_{\nu =2} A_{\nu} F_{\nu}
(r, \Omega), \eqno(5.30)$$
$$O_1 = B_1  \sum_{\nu=2}^N (\nu -1) A_{\nu} F_{\nu} (r, \Omega) \cos
(\varphi_n + \chi_1) + C_1 \cos (\varphi_n + \psi_1), \eqno(5.31)$$
and
$$\eqalign{
O_2 &= \sum_{\nu=2}^N (\nu - 1) A_{\nu} F_{\nu} (r, \Omega) \biggl[ B_2
\cos (2 \varphi_n + \chi_2) + B_3 \cr
&\,+ \frac{1}{2} (\nu - 2) B^2_1 \cos^2 (\varphi_n + \chi_1)
 + \frac{1}{2} B_1 i_n \cos (\varphi_n + \chi_1) \biggr] \cr
&\,+ {1\over 2} C_1 i_n \cos (\varphi_n +\psi_1)
+ C_2 \cos (2\varphi_n + \psi_2) + C_3 . \cr} \eqno(5.32)$$
These formulae contain all terms to second order in $\e^{-rT}$ and to all
non-negative (fractional) powers of $(\frac{\Omega^{\eta}}{r})$.

In the derivation of the map, equations (5.15) and (5.29)--(5.32),
we used two unrealistic simplifications:
\item{a)} The equation of motion of the $y$--variable from equation
(4.8):
$$\ddot y+2r\dot y +(y+x_0)^q -x^q_0 -[(y+x_0)^\ell -x_0^\ell]
\sin^p \Omega t=0 \eqno(5.33)$$
\itemcon cannot be handled by the adiabatic approximation near $t=0$ and
$t=T$ because there $\dot x_0/x_0$ is {\it not} small (see
Figure 2).
\item{b)} In the same regions, $(\sin \Omega t)^\delta$ {\it does not}
yield a reasonable approximation to $x_0(t)$, therefore (5.23) is not
correct in this region.

\noindent Nevertheless, the correction for both effects is completely
absorbed into a proper choice of the constants $I_a$, $\varphi_a$,
$K$ and $P$; {\it the map equations remain unchanged.}

We shall describe here the principal idea only; technical details
will be given in a forthcoming paper.

In the vicinity of $\tau =0$, we use the scaling of Section 4 and
solve
$$\ddot \xi +\xi^q -\xi^\ell \tau^p =0 \;, \eqno(5.34)$$
as a good approximation for the motion within the asymptotically
interesting range of parameters $r$, $\Omega$. Actually, this
``vicinity'' becomes arbitrarily large in the asymptotic limit and we
may consider initial and final times $-\tau_- \gg 1$ and $\tau_+ \gg 1$
large enough that at these times the motion of $\xi (\tau) -
\xi_0 (\tau)$ has become harmonic and adiabatic.

For initial conditions at a {\it large negative} time $\tau_-$
(observe the specular reflection --- we consider model $A$), we use
$$\xi (\tau_-) =-\xi_0 (-\tau_-) +\biggl( {I_a\over 2\omega_0 (\tau_-)}
\biggr)^{1/2} \biggl( u_1 e^{i\Delta \varphi_-} +u_2 e^{-i\Delta
\varphi_-} \biggr) ,\eqno(5.35)$$
$$\dot \xi (\tau_-) =\dot \xi_0 (-\tau_-) +i\biggl( {\omega_0(\tau_-)
I_a\over 2} \biggr)^{1/2} \biggl( u_1 e^{i\Delta \varphi_-}
-u_2 e^{-i\Delta \varphi_-} \biggr) .\eqno(5.36)$$
Here
$$\omega_0^2 (\tau) = (q-\ell) \mid \tau \mid^{(q-1)\delta} \eqno(5.37)$$
and $\Delta \varphi_- (\tau_-)$ will be defined below. At large
positive times $\tau_+$ the solution has the form
$$\xi (\tau_+) =\xi_0 (\tau_+) -\bigl( {2I_+\over \omega_0 (\tau_+)}
\bigr)^{1/2} \cos (\varphi_a +\Delta \varphi_+ (\tau_+)) , \eqno(5.38)$$
$$\dot \xi (\tau_+) = \dot \xi_0 (\tau_+) +(2\omega_0 (\tau_+) I_+)^{1/2}
\sin (\varphi_a +\Delta \varphi_+ (\tau_+)) .\eqno(5.39)$$
For $u_1=u_2=0$, we put
$$I_+=I_a \quad {\rm and} \quad \Delta \varphi_+ = \int_0^{\tau_+}
\omega_a (I_a, \tau) d\tau , \eqno(5.40)$$
thus defining $I_a$ and $\varphi_a$. We may now insert $I_a$ into the
initial conditions (5.35) and (5.36) and put
$$I_-=I_a\cdot \mid u_1\mid^2 \quad {\rm and} \quad \Delta \varphi_- =
-\int_0^{\mid \tau_- \mid} \omega_a (I_-, \tau) d\tau .\eqno(5.41)$$
For sufficiently small $u_1=u_2^*$ we obtain then at $\tau_+$
$$I_+=I_a +\delta I \quad {\rm and} \quad \Delta \varphi_+ =
\delta \varphi + \int_0^{\tau_+} \omega_a (I_+, \tau) d\tau \eqno(5.42)$$
with quantities $\delta I$ and $\delta \varphi$ which are (asymptotically
in $\tau_+$) independent of $\tau_+$ and which have the expansion
(5.8) and (5.9) in powers of $u_1$ and $u_2$.

This construction ensures that all effects of the transition through
the nonadiabatic and nonharmonic region near $\tau =0$ as well as the
effect of $\omega -\omega_a \not= 0$ in this region are fully
absorbed in the well defined limiting quantities $I_a$, $\varphi_a$,
$\delta I$, $\delta \varphi$ and the expansion coefficients of the
latter in powers of $u_\nu$. These quantities are accessible only
numerically.

This completes the derivation of the map.

\section{Summary and conclusions}

We have shown, that the bifurcation behavior of the generalized Duffing
equation
$$\ddot{x} + 2 r \dot{x} + x^q = x^{\ell} \sin^p \Omega t \eqno(6.1)$$
can asymptotically be understood from a {\it one dimensional}
(angle-) Poincar\'e map
$$\varphi_{n+1} = \alpha + \beta \cos \varphi_n. \eqno(6.2)$$
Two aspects of the ``asymptotic'' limit are {\it essential} assumptions
with
physical contents; a third one is more technical.

Firstly, we assumed the validity of the adiabatic approximation which is
sufficiently well fulfilled if $r$ and $\Omega$ are of order
$10^{-1}$ or less.
Secondly, we argued that the Poincar\'e map for one half cycle is
essentially
one dimensional and only regards the angle $\varphi$ if $\e^{-r T}$ is
small enough; $rT \ge 2$ suffices. The obvious reason for the latter
being, that the action contracts by $\e^{- 2 rT}$ during each half cycle.
Both conditions are indispensible to obtain the map (6.2), although
we believe that
qualitatively the structures of the ``tongues'' in the Duffing's equations'
bifurcation diagram remain the same even outside this asymptotic region.

Thirdly, we have derived analytic expressions for the constants $\alpha$
and $\beta$ in terms of $r, \Omega$ and numerical constants, which are
summarized below. For these expressions to be nearly correct, one needs
to go further into asymptopia.
The factor $\e^{-rT}$ roughly estimates the {\it relative} error in the
$\beta \cos \varphi_n$ term of (6.2); including a deviation from the pure
cosine behavior as one sees in (5.31).
In addition, the quantity $\Omega^{\eta}/r$ must be
rather small to justify
the neglect of higher terms in the $\nu$--sums.

Thus there exists a wide region in $r$-$\Omega$ parameter space, where
a one dimensional $\varphi$--map with first and second Fourier components
is a very good approximation
to the Poincar\'e map for (6.1), but where analytic expressions for
the coefficients are not easily available. Actually, for applications,
this is the more interesting region. Although following Section~5 one
may derive
corrections to the formulae given below, it may be
more practical to simply determine a one dimensional map by numerical
integration of (6.1) for one half period $T$. This will be described in
the forthcoming paper.
The asymptotic $aav$'s are best initialized and compared with map values
at the point $t=T/2$.

Collecting all the bits and pieces from the previous Sections, we have
obtained the following asymptotic expressions for the parameters of the map
(6.2):
$$\alpha = A_0 + A_1 T + \sum_{\nu =2}^N A_\nu  F_\nu  (r, \Omega),
 \eqno(6.3)$$
$$\beta = \e^{- rT} [C + B_1 \sum_{\nu  =2}^N (\nu  - 1) A_\nu  F_\nu
  ( r, \Omega)], \eqno(6.4)$$
where $A_0=\varphi_a + \chi_1, B_1$ and $C=C_1 \cos (\psi_1- \chi_1)$
 relate to the
numerical constants defined in Section~5.
We have shifted all angles by $\chi_1$ to obtain the pure cosine
behaviour in (6.2).
The omitted term $C_1 \sin (\psi_1 - \chi_1) \sin \varphi_n$ contributes
 a change
in amplitude and a phase shift which are of order
$(\Omega^{\eta}/r)^{-\kappa_2}$, but its neglect is good only in the
extreme asymptotic limit.

The factors are
$$A_1 = {c_1\over \sqrt{\pi}}
\frac{\Gamma (\frac{\kappa_1}{2})}{\Gamma(\frac{\kappa_1 +1}{2})},\qquad
A_\nu = \nu  c_\nu  I_a^{\nu -1}, \eqno(6.5)$$
$$F_\nu  = 2^{- \kappa_\nu } \Omega^{(\eta -1) \kappa_\nu}
e^{-\pi (\nu -1) r/\Omega} {2\pi \Gamma (\kappa_\nu) \over
\Gamma (1-\delta_\nu +i{(\nu -1) r\over \Omega}) \Gamma (1-\delta_\nu
-i{(\nu -1)r\over \Omega})}$$
$$\approx \frac{\Gamma (\kappa_\nu )}
{(\nu -1)^{\kappa_\nu }} \left(
\frac{\Omega^{\eta}}{2 r}\right)^{\kappa_\nu }, \eqno(6.6)$$
which contain the fourth numerical constant $I_a$ and the
Taylor coefficients $n c_n$ from (2.34).

The exponents were
$$\delta = \frac{p}{q - \ell} < \frac{1}{2}, \qquad\eta=
\frac{(q -1) \delta}{2 + (q -1) \delta}, \eqno(6.7)$$
$$\delta_\nu  = \frac{\delta}{2} [ \frac{\nu }{2} (q +3) - (q + 1) ] ,
 \qquad \kappa_\nu
= 1 - 2 \delta_\nu  \eqno(6.8)$$
and $N$ is to be determined by $\kappa_N > 0, \kappa_{N+1}<0$.

The foregoing considerations are in principle easily extended to the
 general case
of a potential $V(x,\Omega t)$ as described in Section~1. In the
asymptotic region of small $\Omega$ and $r$, the system oscillates
most of the time harmonically around the slowly moving minimum of
$V$, given by Equation (1.3). This motion determines the leading term of
the angle increment as $A_1T$ with
$$A_1={1\over \pi} \int^{s_n+1}_{s_n} [V''(x_0(\tau), \tau)]^{1\over 2}
\d\tau ,\eqno(6.9)$$
where $s_n=\Omega t_n$ and $s_{n+1} =\Omega t_{n+1}$ are two
subsequent time instants at which $V''(x_0(s),s)=0$.
They span a smooth ``period'' between two kicks.
Everything else happens in a narrow region around the $t_n$, where $V$
 may be
approximated by its lowest order terms in $x$ and $t$. Let us take
$t_n=0$ and $x_0(t_n)=0$ for simplicity of notation. Then, $V'(0,0)=0$,
$V''(0,0)=0$, and we may write, to lowest order in $x$ and $t$, with
proper scaling of time and length
$$V(x,\Omega t) \sim {1\over q+1} \mid \xi \mid^{q+1}
-{1\over \ell +1}{\rm sgn} (\xi) \mid \xi \mid^{\ell +1}
\sigma_\tau  \mid \tau \mid^p
 \eqno(6.10)$$       i.e.
$$V'(x,\Omega t) \sim {\rm sgn} (\xi) \mid \xi \mid^q -
\sigma_{\tau} \mid \xi \mid^\ell \mid \tau \mid^p .\eqno(6.11)$$
One has to assume $q>1$ and $q>\ell$ lest the assumptions on $V''$ and
on $x_0$, respectively, are violated. (The condition $2p<(q-\ell)$
is {\it not necessarily} fulfilled. In that case, our theory is not
applicable!)
The sign function \break $\sigma_\tau
=(\tau /\mid \tau \mid )^\nu$, $\nu =1$ or 0, allows for the two models
$A$ or $B$ respectively
for which $x_0(t)$ crosses or touches the $t$-axis.
Except for additional
numerical factors which enter through the foregoing scaling, we obtain
the map (6.2) from $\varphi (t_n)$ to $\varphi (t_{n+1})$, where
the factors now have to be determined by considering
the motion through two previous ``periods''.
Thus, if several zeros at different time instances occur
in $V''$ during one period $2 \pi/\Omega$, one had to introduce one map with
possibly different
parameters for each $t_n$. The bifurcation diagram is then determined
by the finite sequence of maps which describes the alternation between
adiabatic motion and kicks which the system experiences during one full
cycle. In the models of Section~4, we exploited the symmetry between
both half periods to reduce all effects to one single map (6.2).

The original Duffing equation
$$\ddot X +2R\dot X +X^3 +X = A \sin \omega \tau \eqno(6.12)$$
does not obviously belong to the class of nonlinear systems we have
considered since $V'' >0$ throughout in (6.12). Its bifurcation
diagram is often considered [2] in the $A$-$\omega$ plane for fixed
$R$ and large $A$ and has features very similar to the one
without the linear restoring force. This is explained as follows.

Let $a=A^{1/3} >0$ and rescale
$$x={X\over a}, \quad t=a\tau, \quad \Omega ={\omega \over a}, \quad
r={R\over a},\eqno(6.13)$$        to obtain
$$\ddot x +2r\dot x +x^3 +\bigl( {r\over R} \bigr)^2 x = \sin
\Omega t.\eqno(6.14)$$
The asymptotic behavior in the $r$-$\Omega$ parameter plane with $R$
held fixed is now described by our theory with $q=3$,
$\ell =0$, $p=1$. The term $(r/R)^2 x$ is negligible if $(r
\Omega^{-\gamma}) \ll R$; we had neglected a similar term in
Section~2.

An interesting question arises how to understand bifurcation diagrams
for systems, the potentials of which do not obey all assumptions made
in Section~1.

One could consider systems where $V''>0$ remains finite
asymptotically in $r$ and $\Omega$ for all times $t$.
Our adiabatic approximation scheme breaks down in
this case as it yields $I(t)\sim \e^{-2rt} I(0)$ forever.
It is conceivable that such systems have no bifurcations at all in
the asymptotic limit.
Otherwise, one expects that ${r\over \Omega} \simle 1$
is the parameter regime of interest. One could imagine, then, that
some correction to the adiabatic approximation leads to an equation
for slow motion of $I$, which might yield a map upon completion of
a cycle.

Another possibility for $V$ is that $V''=0$ does occur, but that $V$ has
a multiple well structure, at least during some time intervals
(for instance, the periodically driven Josephson junction, a system of
great interest). In that case, the adiabatic approximation applies during
most of the time, but switching between different wells will yield a wealth
of new phenomena, which are likely not amenable to the simple type of
analysis we have used here.

Correspondence with W. Lauterborn, Darmstadt, on the manuscript is gratefully
acknowledged.

\references
\numrefjl{[1]}{Duffing G 1918}{ Erzwungene Schwingungen bei ver\"anderlicher
Eigenfrequenz und ihre technische Bedeutung}{ (Braunschweig: Vieweg)}
\numrefjl{}{Ueda Y 1980}{ Steady motions exhibited by Duffing equation: a
picture book of regular and chaotic motions}{  New Approaches to Nonlinear
Problems in Dynamics (P J Holmes, Ed)}{ SIAM 1980 331-322}
\numrefjl{}{Guckenheimer I, Holmes P 1983}{ Nonlinear Oscillations,
Dynamical Systems and Bifurcations of Vector Fields 82-92}{(New York,
Berlin, Heidelberg, Tokyo: Springer)}
\numrefjl{}{Byatt-Smith J G 1986}{ Regular and Chaotic Solutions of
Duffing's equation for Large Forcing}{ IMA Journal of Applied Mathematics
37 113-154}
\numrefjl{[2]}{Parlitz U, Lauterborn W 1987}{Superstructure in the
bifurcation set of the Duffing equation}{ Physics Letters 107A 351-355}
\numrefjl{}{Parlitz U, Scheffczyk T, Kurz T, Lauterborn W 1991}{ On
modelling driven oscillators by maps }{Int. I. Bifurcation and Chaos
(To appear)}
\numrefjl{[3]}{Sato S, Sano M, Sawada Y 1983}{ Universal scaling property
in bifurcation structure of Duffing's and of generalized Duffing's
equations}{ Phys. Rev. A 28 1654-1658}
\numrefjl{[4]}{Brown R and Chua L 1990}{ Horseshoes in the twist and
flip map} {Memorandum UCB/ERL M90/80}
\numrefjl{[5]}{Gradsteyn I S, Ryshik I M 1965}{ Table of Integrals, Series
and Products}{ (New York, London: Academic Press)}

\figures

\figcaption{Step function driving as kick mechanism.}

\figcaption{Reference solution with jumps.}

\figcaption{Motion of the potential and of the reference solution.}

\bye